\documentclass[namedreferences,hyperref,optionalrh]{spr-sola}
\usepackage{graphicx}        
\usepackage{color}           

\renewcommand{\vec}[1]{{\mathbfit #1}}

\newcommand{\apj}{{\it Astrophys. J.}}

\newcommand{\solphys}{{\it Sol. Phys.}}

\chardef\us=`\_

\usepackage{lineno}
\begin{document}

\begin{frontmatter}
\title{\textbf{Comparison of Relative Magnetic Helicity Flux Calculation Results Based on the Line-of-Sight Magnetograms of ASO-S/FMG and SDO/HMI}}

\author[addressref={aff1,aff2,aff3},email={yangshb@nao.cas.cn}]{\inits{S.}\fnm{Shangbin}~\snm{Yang}\orcid{0000-0002-2967-4522}}
\author[addressref={aff1,aff3},email={lius@nao.cas.cn}]{\inits{S.}\fnm{Suo}~\snm{Liu}}
\author[addressref={aff1,aff2,aff3},email={sjt@nao.cas.cn}]{\inits{J. T.}\fnm{Jiangtao}~\snm{Su}}
\author[addressref={aff1,aff2,aff3},email={dyy@nao.cas.cn}]{\inits{Y. Y.}\fnm{Yuanyong}~\snm{Deng}}
\address[id=aff1]{Huairou Solar Observing Station, National
Astronomical Observatories, Chinese Academy of Sciences, 100101
Beijing, China}
\address[id=aff2]{University of Chinese Academy of Sciences, 100049 Beijing, China}
\address[id=aff3]{Key Laboratory of Solar Activity and Space Weather, 100190 Beijing, China}

\runningauthor{Yang et al.}
\runningtitle{Yang et al. 2024}

\begin{abstract}
Magnetic helicity is a key geometrical parameter to describe the structure and evolution of
solar coronal magnetic fields. The accumulation of magnetic helicity is correlated with the
non-potential magnetic field energy, which is released in the solar eruptions. Moreover, the relative magnetic
helicity fluxes can be estimated only relying on the line-of-sight magnetic field (e. g. \citeauthor{Demo03} \solphys{} \textbf{215}, 203, \citeyear{Demo03}). The payload Full-disk MagnetoGraph (FMG) on the Advanced
Space-based Solar Observatory (ASO-S) currently has been supplying the continuous evolution of line-of-sight magnetograms for the solar active regions, which can be used to estimate the magnetic helicity flux.
In this study, we useeight hours line-of-sight magnetograms of NOAA 13273, when at which the Sun-Earth direction speed of the satellite is zero to avoid the oscillation of the magnetic field caused by the Doppler effect on polarization measurements.
We obtain the helicity flux by applying Fast Fourier Transforms (FFT) and local correlation tracking (LCT) methods
to obtain the horizontal vector potential field and the motions of the ine-sf-sight polarities.
We also compare the helicity flux derived using data from the Heliosesmic and Magnetic Imager (HMI) on board the Solar Dynamics Observatory (SDO) and the same method. It is found that the flux has the same sign and the correlation between measurements is 0.98. The difference of the absolute magnetic helicity normalized to themagnetic flux is less than 4\%. This comparison demonstrates the reliability of ASO-S/FMG data and that  it can be reliably used in future studies.

\end{abstract}
\end{frontmatter}

\section{Introduction}
     \label{S-Introduction}

It is well known that solar flares and coronal mass ejections
(CMEs) are closely related to the sudden explosive release of stored
free magnetic energy in the corona \citep[see, e.g.][]{Forbes2002}).
Unfortunately, magnetic fields can be observed only in the
photosphere and only few indirect measurements exist to determine the
magnetic field in the corona
\citep[e.g.][]{Lin04}. Although linear and nonlinear mathematical
approaches are used to extrapolate force-free magnetic fields from
the photosphere into the corona it is still difficult to estimate
the free magnetic energy contained in the corona and the conditions
of its release \citep{Sch06}.

Magnetic helicity is a quantity describing the twist, writhe, and
torsion of magnetic field lines and magnetic configurations
\citep{Berger99}. The concept of magnetic helicity has
successfully been applied to characterize solar coronal processes,
for a recent review about observations and computation of the photospheric
magnetic helicity see \citet{Dem09}. \citet{ZhangM06,ZhangM08} conjectured that there is an upper limit
of the total magnetic helicity that a force-free magnetic field can
contain before it erupts. It also has been found that the temporal
evolution of the relative magnetic helicity is closely related to
the evolution of the magnetic energy content of the corona
\citep{Santos11,Skala15,Yang13}. For a better understanding of the energy storage and release in the
corona it is, therefore, essential to investigate the magnetic
helicity accumulation. \citet{Park08} investigated the variation of the
magnetic helicity for the 11 X-class flares that occurred in seven active regions (ARs). They found that each of these major flares was preceded by a
significant helicity accumulation at a nearly constant rate before the flares. \citet{ZhangM06,ZhangM08} also
found that there is a sharp jump of magnetic helicity change rate
closely related to the solar eruption.
\citet{Nindos04} investigated 133 events and showed that the amount of the stored pre-flare coronal helicity could determine if a flare is eruptive or confined.
\citet{Tzio12} analyzed the free magnetic energy and relative magnetic helicity budgets in
162 vector magnetograms and found that a statistically robust, monotonic correlation between the free magnetic energy and the relative magnetic helicity which indicates that the relative magnetic helicity is an essential ingredient for major solar eruptions. Observations and simulations also show a threshold for current-carrying parts of relative magnetic helicity \citep{Price19,Gupta21}. \citet{Lio22,Lio23} also discussed the role of left- and right-handed accumulated helicity and found that the excess budget is cotemporal with flare peaks. Both results discussed above also show that
there is correlation between the evolution of magnetic helicity and solar
eruptions. Moreover, the observation of magnetic helicity patterns over long periods could be used to analyze the so-called helicity hemisphere rule \citep{Bao98,Zhang10}, as well as to
constrain the solar dynamo from the point of view of magnetic helicity
conservation over large scales and modulation of the solar cycle strength \citep{Yang20}. A comprehensive review about the investigation of magnetic helicity in geophysics and astrophysics can be found in \citet{Kuzanyan24} and references therein.

As the first launched Chinese solar satellite, Advanced Space-based Solar Observatory (ASO-S) is focused  on observing solar magnetic fields, solar flares
, and coronal mass ejections (CMEs) in order to study their relationships, which are key scientific questions in modern solar physics \citep{Gan23}.Three payloads
are therefore deployed on the ASO-S: the Full-disk vector MagnetoGraph \citep[FMG]{Deng19}, the Lyman-$\alpha$ Solar Telescope
\cite[LST]{Feng19}, and the Hard X-ray Imager \citep[HXI]{Su19}.
In order to study the correlation between the accumulated
magnetic helicity and coronal dynamics we analyzed the helicity evolution using ASO-S/FMG data. In this study, we use line-of-sight ASO-S/FMG magnetograms to estimate
the magnetic helicity flux and compare the results with those derived using
the Heliosiesmic and Magnetic Imager (HMI) data on board the Solar Dynamics Observatory (SDO); ; this allows us to verify the feasibility of FMG data for helicity studies.In Sec.~\ref{S-DM}, we introduce the data processing and magnetic helicity flux estimation method. In Sec.~\ref{S-Results}, we presented the comparison of magnetic field, velocity flows, magnetic helicity density maps and magnetic helicity accumulation computations between ASO-S/FMG and SDO/HMI data. A summary is presented in Sec.~\ref{S-Summary}.

\section{Data and Method} 
  \label{S-DM}

In this article, we use the data from ASO-S/FMG and SDO/HMI for the comparison of magnetic helicity.  The aperture of FMG is 14 cm. A CMOS camera with 4096 × 4096 pixels is used. FMG observes polarized
Stokes images using a liquid crystal variable retarder (Hou et al., 2020). The
routine observations of the FMG are taken at one wavelength position of the Fe {\sc i}
5324.19 \AA~and a linear calibration method is adopted. The longitudinal magnetic
field sensitivity is 15 G (normal mode). The temporal resolution is 30 second
(single component) and 2 minutes (vector magnetograms) under normal mode.
The pixel size is about $0.5^{\prime\prime}$ and the spatial resolution of FMG for the flight
model (FM) is $1.04^{\prime\prime}$ (Deng et al., 2019; Su et al., 2019; Gan et al., 2023).
HMI is an instrument designed to study oscillations and the magnetic field at the solar photosphere. HMI is one of three instruments on the SDO. It observes the Sun nearly continuously and takes one terabyte of data per day. HMI observes the full solar disk at 6173 Å with a resolution of $1^{\prime\prime}$ \citep{Schou12}.

To calculate magnetic helicity fluxes
across the photosphere \textbf{S} , we use the following equation:
\begin{equation}
{\frac{dH_{R}}{dt}=-2\int(\vec{A}_{p}\cdot\vec{U}){B_n}\vec{dS},}
\label{eq:Hrate}
\end{equation}
where $\vec{U}$ denotes the horizontal velocity flows by
that are computed using local correlation tracking (LCT). The vector potential $\vec{A}_{p}$ is obtained using the fast Fourier transform (FFT) applied to  the normal components of the photospheric magnetic field $B_n$ \citep{Chae01}. $B_n$ is estimated by multiplying the magnetic field
line-of-sight component by $1/cos\psi$ where $\psi$ is the
heliocentric angle of the region \citep{Liu06,Yang09a,Yang09b}. According to the derivation of
\citet{Demo03}, 
Equation 1 includes the helicity
flux from the emerging of magnetic flux and
the motions of the magnetic field elements at the solar photosphere.
 Such approach has been widely used to calculate relative magnetic helicity flux across the solar photosphere to the corona \citep{Green02,Nindos02,Moon02,Nindos03}. 
 Currently, we cannot obtain the continuous vector magnetograms of solar active regions from the FMG data. The above equation indicates that we can evaluate the magnetic helicity accumulation in the solar corona evolution only based on a series of line-of-sight magnetic fields.

In this tudy, we choose active region (AR) NOAA 13273 when it crossed the solar disk center on 10 Arpil from 00:00 to 08:00 UTC. The longitude of this active region in the chosen time interval is less than E25 that reduce the projection effect.When processing the data, we have considered:

1) The resolution and spatial alignments.

Since the optical diffraction limit of FMG and HMI are both close to $1^{\prime\prime}$, we use the cubic convolution interpolation method to rescale the data to $1^{\prime\prime}$  pixel resolution. Then, we cropped the same region to 256 x 256.

2) The cadence and time alignments.

The normal mode of ASO-S/FMG is to observe the polarization signal at the fixed off-band of Fe {\sc i} 5324.19 \AA. The Doppler shift caused by the satellite orbit change the working wavelength position of Lyot narrow-band filter of FMG and take the oscillation of measured polarization signals. It will produce an uncertainty of solar magnetic field measurements. To avoid this problem, we use the orbit speed record in the header of FMG fits file, and apply a cubic convolution interpolation method to obtain the processed FMG data in which the orbit Sun-satellite direction speed is zero. Then, we choose the HMI magnetic field data at the closet time and also apply the cubic convolution interpolation method to obtain the processed HMI data at the same time. Note that the new observing times of FMG and HMI data have been changed after the interpolation. The cadence after time alignment is 52 minutes.

3) The tracking parameter setting.

Physically significant transverse velocities of the photospheric elements
are usually smaller than 1.5 km s$^{-1}$ \citep{Chae01}. Hence the
possible shift between two consequent magnetograms is less than 8
pixels in the estimated time cadence. For the LCT method, we choose $8''$ for the FWHM (full
width at half-maximum) of the apodizing function . To reduce the
noise, we set the horizontal velocity to zero in regions where the
magnetic field is small ${(< 10 G)}$. In order to better track the
emerging regions and to exclude the effect of relative quiet regions
outside the emergence sites, we set the horizontal velocity to zero
in regions of a weak cross-correlation (${<0.9}$) between two
magnetograms.

After the above data processing, the final sizes of FMG and HMI
data cubes is 256 x 256 x 10. The first two dimensions are the solar X and Y coordinate directions. The third dimension is for the new interpolation observing time. In the next section, we will calculate the helicity flux and compare the results.



\section{Results}
  \label{S-Results}

\begin{figure}    
\centerline{\includegraphics[width=1.0\textwidth,clip=]{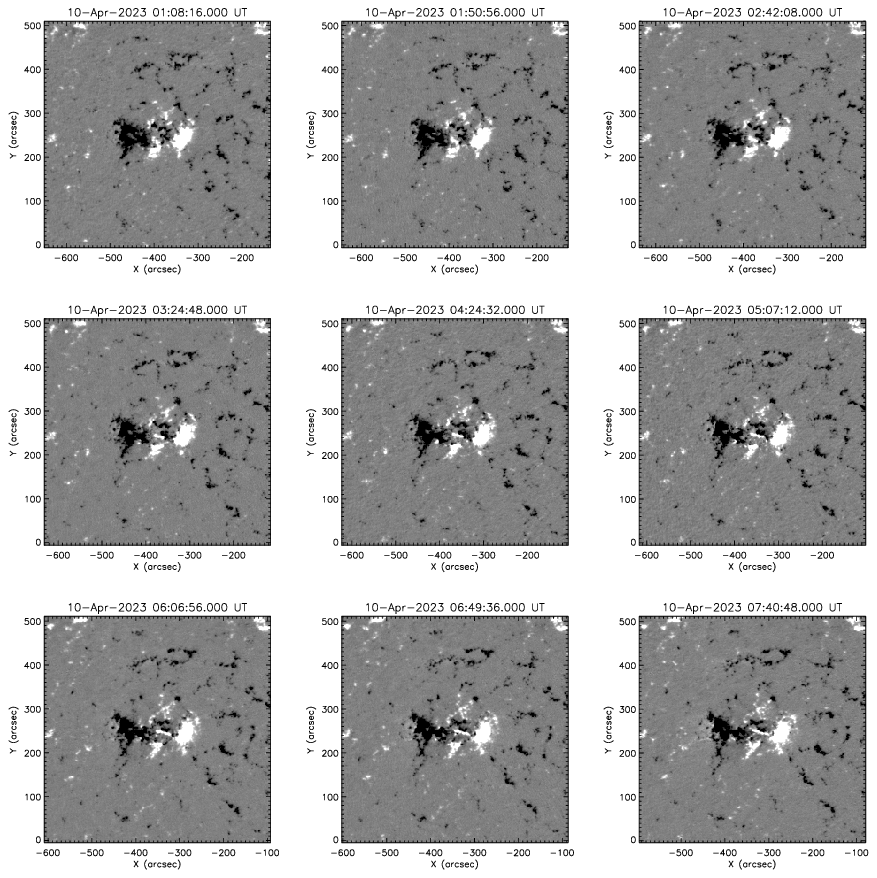}}
\small
\caption{Evolution of the line-of-sight magnetic field of AR NOAA13273 from ASO-S/FMG. It is scaled to [-250G, 250G]. }
\label{fig:LOSB_FMG}
\end{figure}

\begin{figure}    
\centerline{\includegraphics[width=1.0\textwidth,clip=]{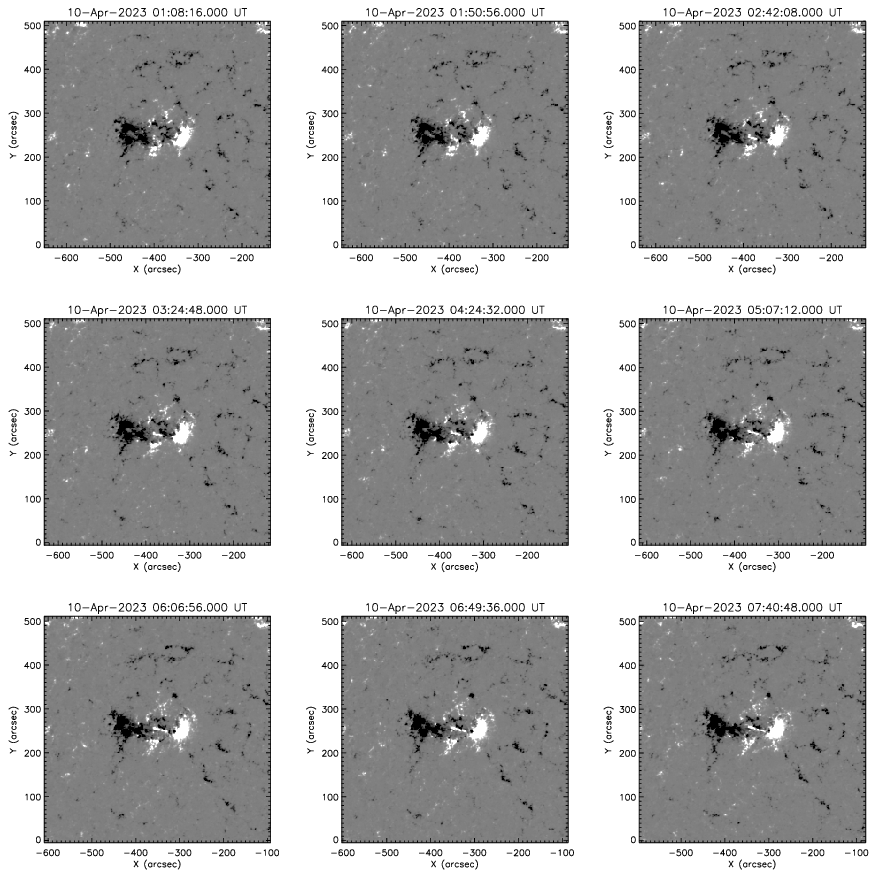}}
\small
\caption{Evolution of the line-of-sight magnetic field of AR NOAA13273 from SDO/HMI.It is scaled to [-250G, 250G].}
\label{fig:LOSB_HMI}
\end{figure}

\begin{figure}    
\centerline{\includegraphics[width=1.0\textwidth,clip=]{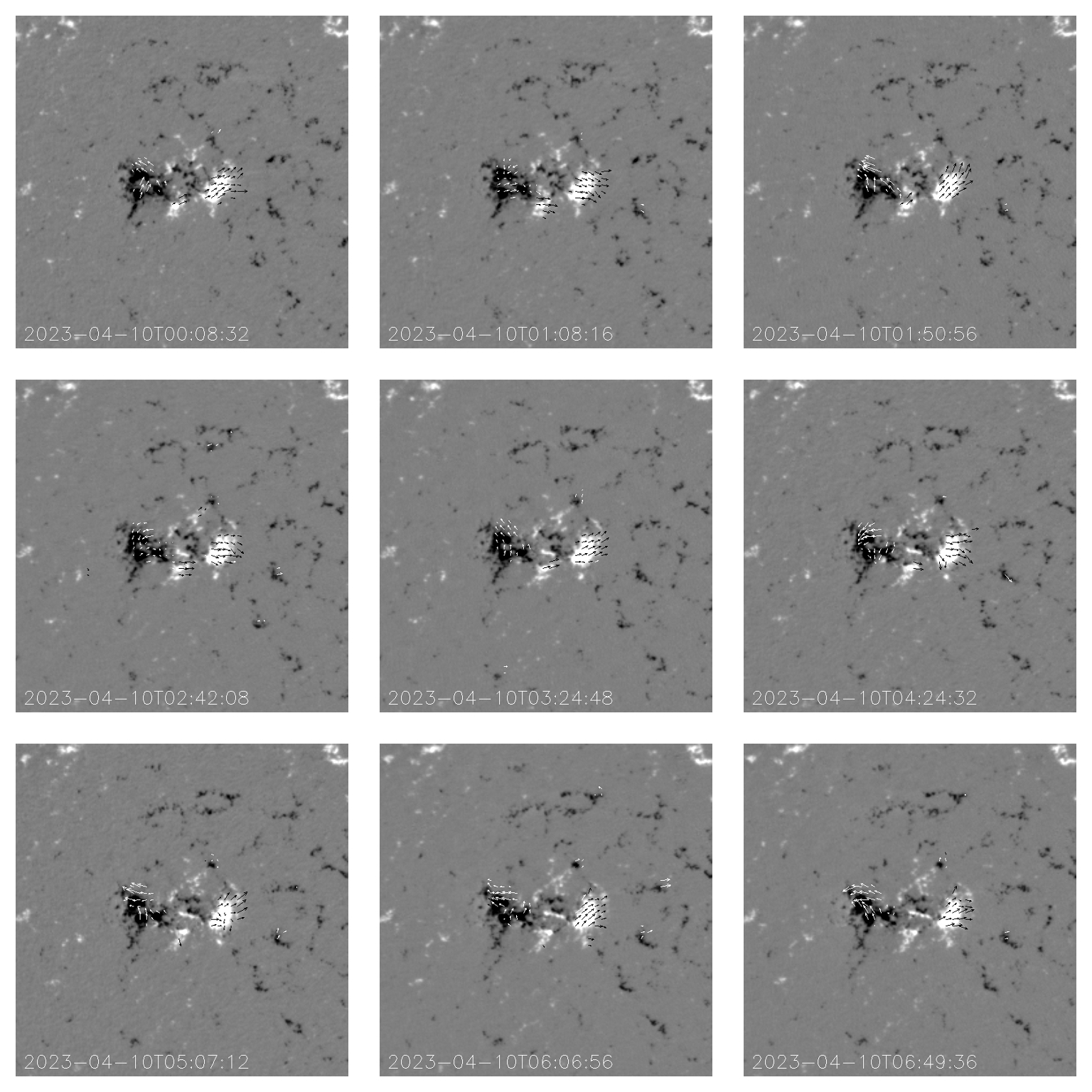}}
\small
\caption{Evolution of the horizontal velocity of AR NOAA13273 from ASO-S/FMG. The maximum arrow length measures a velocity of 0.8 km/s.}
\label{fig:Vmap_FMG}
\end{figure}

\begin{figure}    
\centerline{\includegraphics[width=1.0\textwidth,clip=]{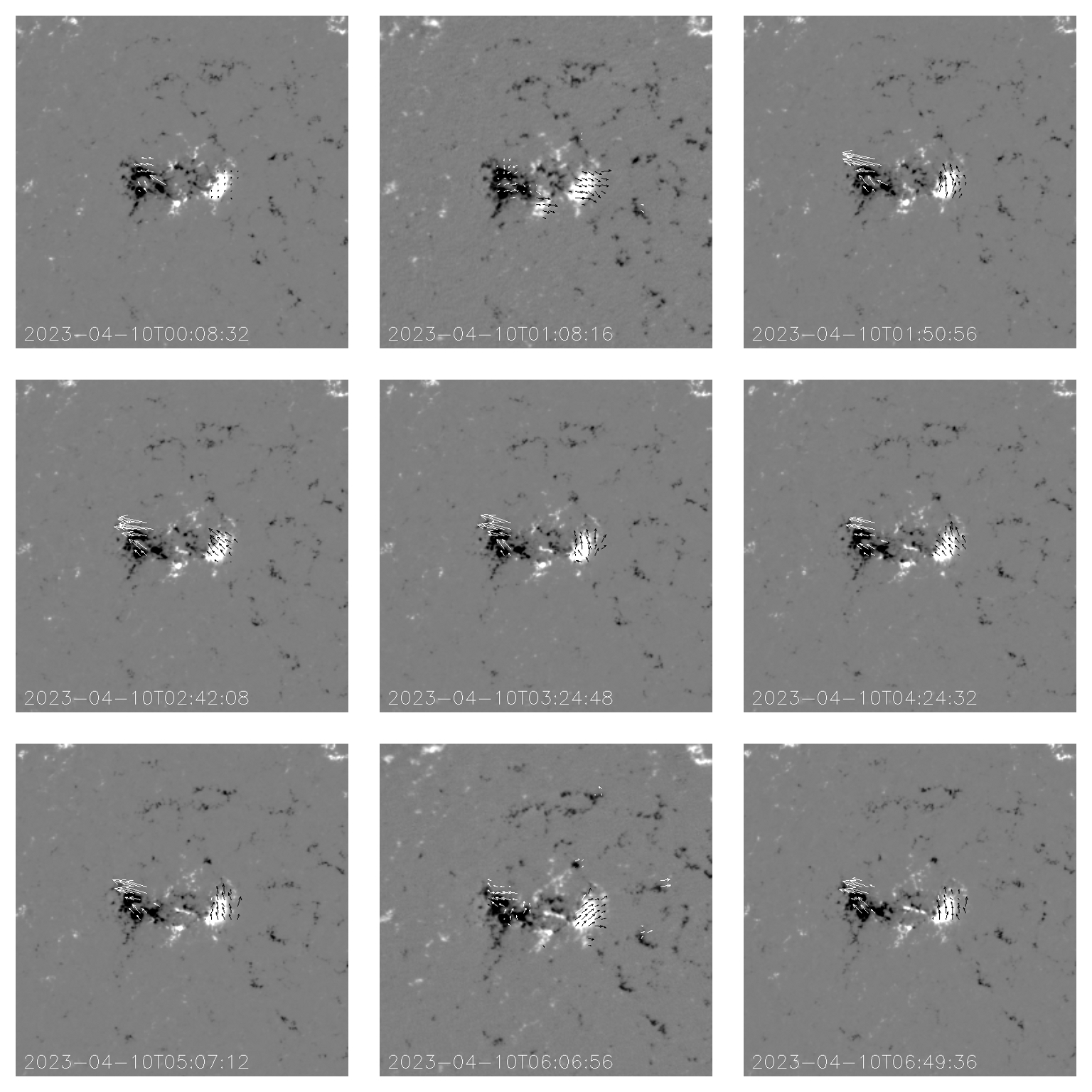}}
\small
\caption{Evolution of the horizontal velocity of AR NOAA13273 from SDO/HMI. The maximum arrow length measures a velocity of 0.8 km/s.}
\label{fig:Vmap_HMI}
\end{figure}

\begin{figure}    
\centerline{\includegraphics[width=1.0\textwidth,clip=]{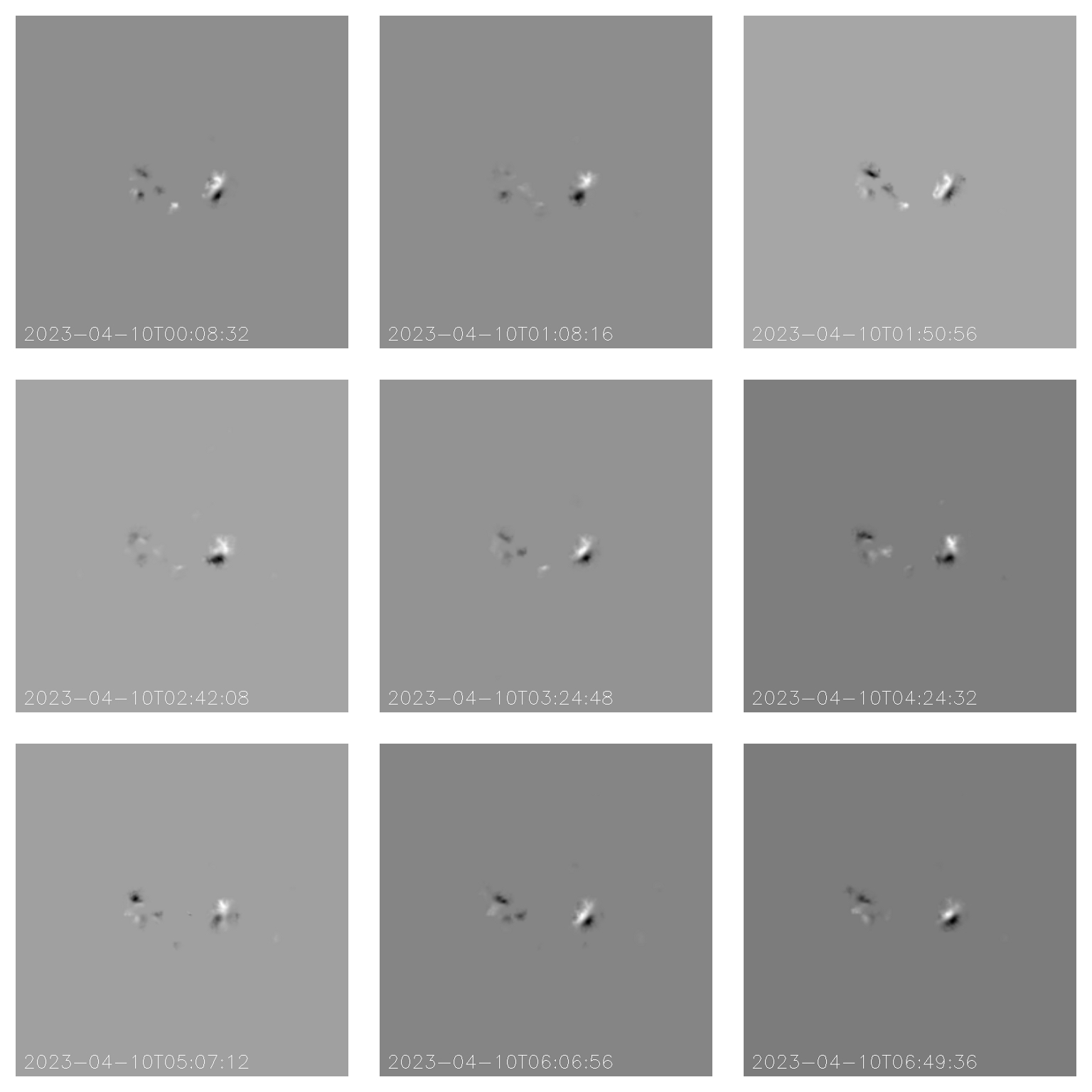}}
\small
\caption{Evolution of the magnetic helicity density map $G_A=-2(A_p  \cdot U)B_n$ of NOAA13273 from ASO-S/FMG.
The white and black colors indicate the positive and negative signs of the density map.
}
\label{fig:Gmap_FMG}
\end{figure}

\begin{figure}    
\centerline{\includegraphics[width=1.0\textwidth,clip=]{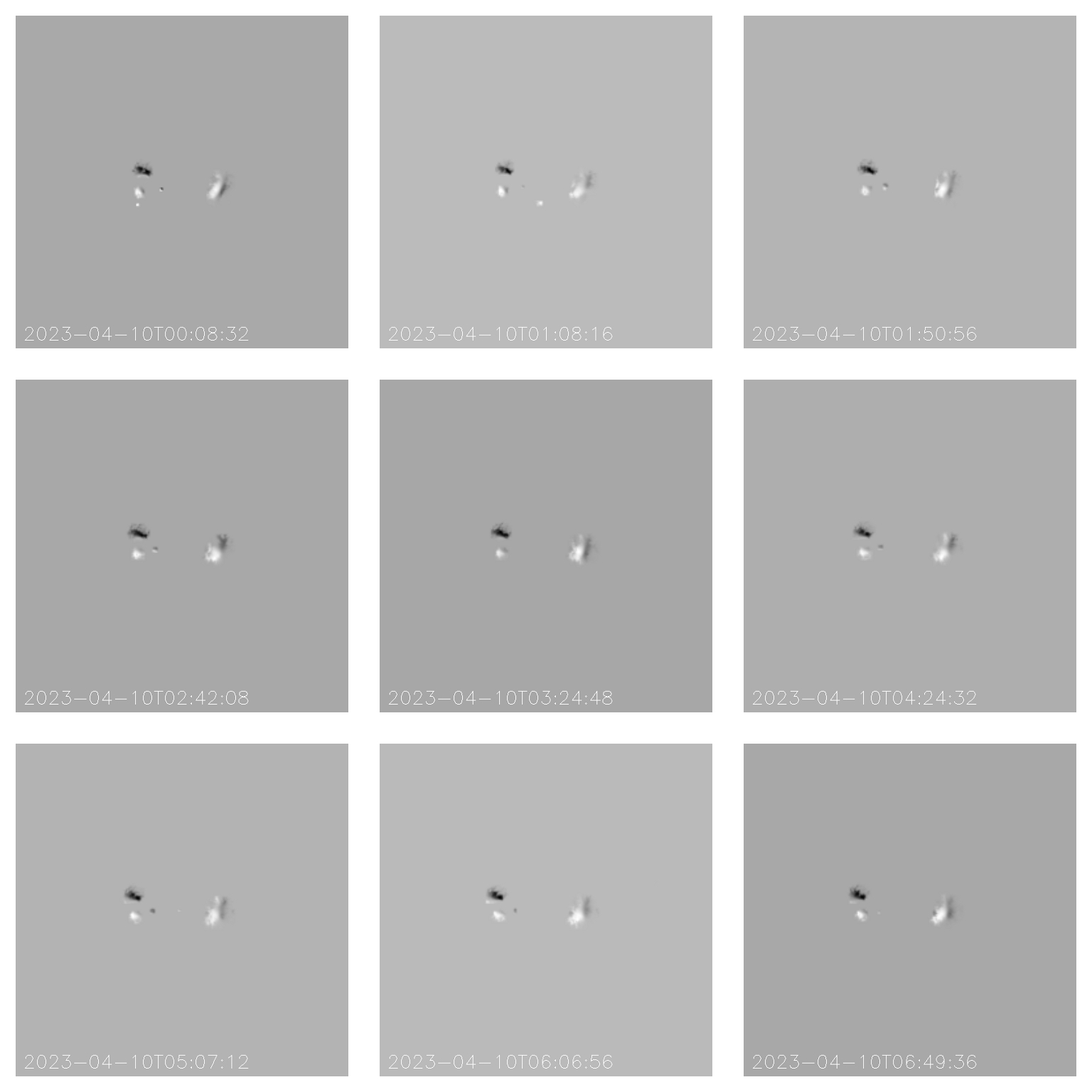}}
\small
\caption{Evolution of the magnetic helicity density map  $G_A=-2(A_p  \cdot U)B_n$ of NOAA13273 from SDO/HMI.
The white and black colors indicate the positive and negative signs of the density map.}
\label{fig:Gmap_HMI}
\end{figure}

\begin{figure}    
\centerline{\includegraphics[width=1.0\textwidth,clip=]{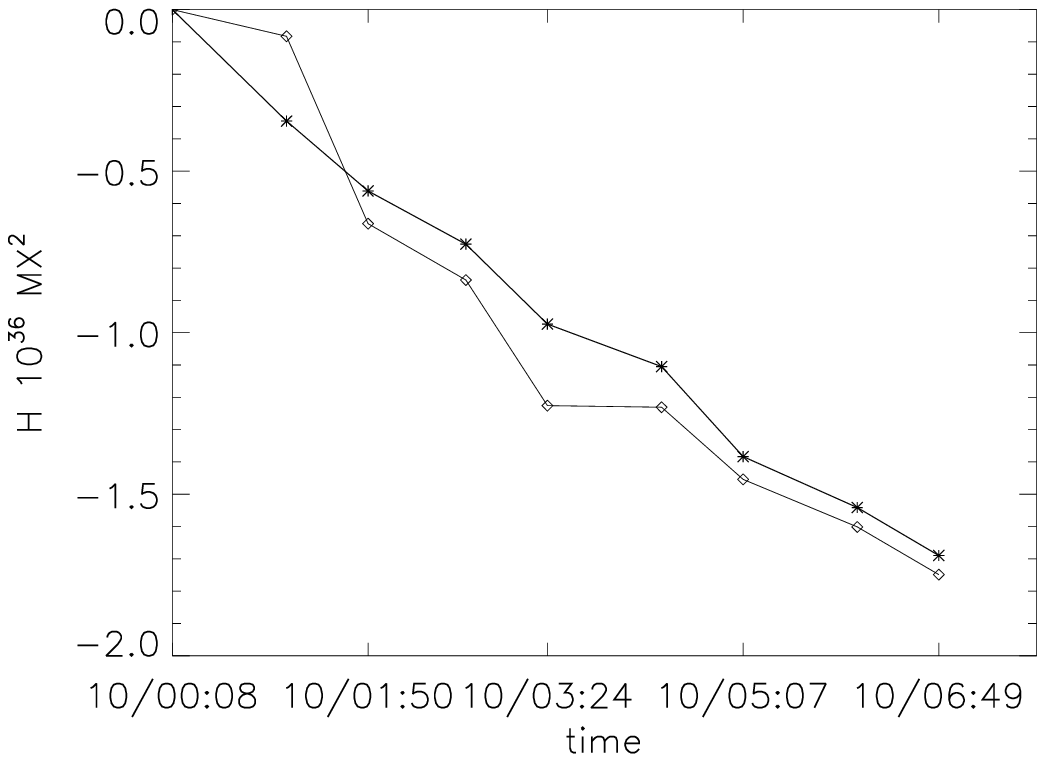}}
\small
\caption{Accumulation of magnetic helicity of NOAA13273 by using FMG data (diamond) and HMI (asterisk).}
\label{fig:HR_comparison}
\end{figure}

Figure~\ref{fig:LOSB_FMG} depicts the evolution of the line-of-sight magnetic field of NOAA 13273 from ASO-S/FMG on 10 April  2023 from 01:00 UTC to 07:40 UTC. Figure~\ref{fig:LOSB_HMI} shows a similar evolution of NOAA 13273 observed by SDO/HMI.
The distribution and evolution of magnetic field  is well consistent and the average for linear Pearson correlation coefficient is 0.93. Figure~\ref{fig:Vmap_FMG} and Figure~\ref{fig:Vmap_HMI} show, respectively, the horizontal velocity 
using the LCT method applied to
 ASO-S/FMG and SDO/HMI. The two figures show consistent motions in general.
We compute the separation between the leading positive polarity and the following negative polarity sunspots. A small emerging positive flux moved toward the leading sunspots and gradually cancelled the negative polarity flux. Figure~\ref{fig:Gmap_FMG}
and Figure~\ref{fig:Gmap_HMI}, respectively, depict the evolution of magnetic helicity flux density map, which is the integration core of Equation~(\ref{eq:Hrate}) defined as $G_A$.
 The two density maps are also consistent in general. The majority of the sign of magnetic helicity density maps is negative. It means that the accumulated magnetic helicity during the calculated time period  is also negative. Note the mixture of positive and negative helicity density in the leading sunspots. This is due to the same trend of the horizontal velocity field and the reverse direction of the vector potential of the magnetic field around the same leading sunspot polarity, creates some fake parity signals into the $G_A$.
  Figure~\ref{fig:HR_comparison} shows the evolution profile of the accumulated magnetic helicity of NOAA 12373 by using ASO-S/FMG (diamond) and SDO/HMI data (asterisk).  The difference of the final accumulated magnetic helicity is only 3.4\%. The correlation of the two magnetic helicity accumulation profiles is 0.98, which reflects the high consistence of magnetic helicity estimation from the ASO-S/FMG and SDO/HMI data.

\section{Summary}
  \label{S-Summary}
Magnetic helicity plays an important role to understand the origin of solar eruptions and even their long-term evolution could be used to constrain the solar dynamo process. Magnetic helicity can quantitatively be calculated using the continuous evolution of the magnetic field, which definitely helpsto push forward the achievementof ASO-S scientific goals. After carefully choosing magnetic field measurements at the optimum moment, Local-Correlation-Tracking parameters under appropriate resolution and cadence with the appropriate projection correction, we can obtain a high trusted calculation of the magnetic helicity flux estimation by comparing with the results based on SDO/HMI data, which is widely used in the previous magnetic helicity flux estimation research. The linear correlation between the profiles is 0.98 and the difference of magnetic helicity accumulation is
less than 4\% for the solar AR in this study. In the future, In the future, we will carry on a large statistically significant study of magnetic helicity evolution and we will analyze its relations with solar flares and CMEs during the solar maximum of Solar Cycle 25.

\begin{acknowledgements}
\small
We would like to thank the referee for carefully reading our manuscript and for giving constructive comments that substantially helped improving the paper. We would also like to thank the editor to significantly improve the language of the manuscript.
\end{acknowledgements}

\begin{itemize}
\item \texttt{fundinginformation}
\end{itemize}

\begin{fundinginformation}
This research is supported by the National Key $R\&D$ Program of China No.
\\ 2022YFF0503800,2021YFA1600503 and 2021YFA1600500; National Natural Science Foundation of China (grants No. 12250005, 12073040, 11973056, 12003051, 11573037, 12073041, 11427901, and 11611530679); by the Strategic Priority Research Program of the Chinese Academy of Sciences (grants No. XDB0560000, XDA15052200, XDB09040200, XDA15010700, and XDA15320102); by the Youth Innovation Promotion Association of CAS (2019059).The authors would  also like to thank the Supercomputing Center of the Chinese Academy of Sciences (SCCAS).
\end{fundinginformation}

\clearpage


\end{document}